\documentclass{emulateapj}

\def\msol{\hbox{\kern 0.20em $M_\odot$}}
\def\lsol{\hbox{\kern 0.20em $L_\odot$}}
\def\rsol{\hbox{\kern 0.20em $R_\odot$}}
\def\sr{\hbox{\kern 0.20em sr}}
\def\srmu{\hbox{\kern 0.20em sr$^{-1}$}}
 
\def\g{\hbox{\kern 0.20em g}}
\def\gmu{\hbox{\kern 0.20em g$^{-1}$}}
\def\kg{\hbox{\kern 0.20em kg}}
\def\pc{\hbox{\kern 0.20em pc}}
 
\def\mum{\hbox{\kern 0.20em $\mu$m}}
\def\mumd{\hbox{\kern 0.20em $\mu$m$^{-2}$}}
\def\cm{\hbox{\kern 0.20em cm}}
\def\m{\hbox{\kern 0.20em m}}
\def\km{\hbox{\kern 0.20em km}}
\def\nm{\hbox{\kern 0.20em nm}}
 
\def\s{\hbox{\kern 0.20em s}}
\def\h{\hbox{\kern 0.20em h}}
\def\sec{\hbox{\kern 0.20em sec}}
\def\min{\hbox {\kern 0.20em min}}
\def\smu{\hbox{\kern 0.20em s$^{-1}$}}
\def\smd{\hbox{\kern 0.20em s$^{-2}$}}
\def\an{\hbox{\kern 0.20em an}}
\def\anmu{\hbox{\kern 0.20em an$^{-1}$}}
\def\deg{\hbox{\kern 0.20em $^{\rm o}$}}
\def\yr{\hbox{\kern 0.20em yr}}
\def\yrmu{\hbox{\kern 0.20em yr$^{-1}$}}
\def\Myr{\hbox{\kern 0.20em Myr}}
\def\Mymu{\hbox{\kern 0.20em Myr$^{-1}$}}
\def\K{\hbox{\kern 0.20em K}}
\def\pcmu{\hbox{\kern 0.20em pc$^{-1}$}}
\def\pcmd{\hbox{\kern 0.20em pc$^{-2}$}}
\def\pcmt{\hbox{\kern 0.20em pc$^{-3}$}}
\def\kms{\hbox{\kern 0.20em km\kern 0.20em s$^{-1}$}}
\def\kmpd{\hbox{\kern 0.20em km$^{2}$}}
\def\kpc{\hbox{\kern 0.20em kpc}}
\def\cms{\hbox{\kern 0.20em cm\kern 0.20em s$^{-1}$}}
\def\erg{\hbox{\kern 0.20em erg}}
\def\ergs{\hbox{\kern 0.20em erg}}
\def\cmpd{\hbox{\kern 0.20em cm$^2$}}
\def\cmmd{\hbox{\kern 0.20em cm$^{-2}$}}
\def\cmms{\hbox{\kern 0.20em cm$^{-6}$}}
\def\cmpt{\hbox{\kern 0.20em cm$^3$}}
\def\cmmt{\hbox{\kern 0.20em cm$^{-3}$}}
\def\mpd{\hbox{\kern 0.20em m$^2$}}
\def\mmd{\hbox{\kern 0.20em m$^{-2}$}}
\def\mpt{\hbox{\kern 0.20em m$^3$}}
\def\mmt{\hbox{\kern 0.20em m$^{-3}$}}
\def\mujy{\hbox{\kern 0.20em $\mu$Jy}}
\def\mjy{\hbox{\kern 0.20em mJy}}
\def\Mj{\hbox{\kern 0.20em MJy}}
\def\jy{\hbox{\kern 0.20em Jy}}
\def\ghz{\hbox{\kern 0.20em GHz}}
\def\srmd{\hbox{\kern 0.20em sr$^{-1}$}}
\def \kms{km~$\rm{s}^{-1}$}

\def \mum{$\mu$m}

\def\G{\hbox{\kern 0.20em G}}

\def\h13cop{\hbox{H$^{13}$CO$^{+}$}}

\def\S+{\hbox{S{\small II}}}

\slugcomment{Submitted ApJLetters on December, 2010}

\shorttitle{The IR symmetry of HH~34}
\shortauthors{Raga et al.}
\def \mum{$\mu$m}

\begin{document}

\title{The jet/counterjet IR symmetry of HH~34 and the size of
the jet formation region}

\author{Raga, A. C. \altaffilmark{1},
Noriega-Crespo, A.\altaffilmark{2},
Lora, V.\altaffilmark{3},
Stapelfeldt, K. R.\altaffilmark{4}
and
Carey, S. J.\altaffilmark{2}}

\altaffiltext{1}{Instituto de Ciencias Nucleares, Universidad Nacional 
Aut\'onoma de M\'exico, Ap. 70-543, 04510 D.F., M\'exico}
\altaffiltext{2}{SPITZER Science Center, California Institute of 
Technology,CA 91125  USA}
\altaffiltext{3}{Astronomisches Rechen-Institut Zentrum f\"ur
Astronomie der Universit\"at Heidelberg, M\"onchhofstr. 12-14
69120 Heidelberg, Germany }
\altaffiltext{4}{Jet propulsion Laboratory, California Institute of 
Technology,MS 183-900, 4800 Oak Grove Drive, Pasadena, CA 91109, USA}

\begin{abstract}
We present a new IRAC, Spitzer IRAC images of the HH~34 outflow. These
are the first images that detect both the knots along the southern
jet and the northern counterjet (the counterjet knots were only detected
previously in a long slit spectrum). This result removes the problem
of the apparent coexistence of a large scale symmetry (at distances of
up to $\sim 1$~pc) and a complete lack of symmetry close to the source
(at distances of $\sim 10^{17}$~cm) for this outflow. We present a
quantitative evaluation of the newly found symmetry between the
HH~34 jet and counterjet, and show that the observed degree
of symmetry implies that the jet production region has
a characteristic size $<2.8$~AU. This is the strongest
constraint yet derived for the size of the region in which
HH jets are produced.
\end{abstract}


\section{Introduction}

HH 34 is one of the HH objects in Herbig's catalogue (Herbig 1974).
It jumped into the limelight with the paper of Reipurth et al. (1986),
who showed that the HH object (HH 34S) had a bow shaped morphology,
and a jet-like association of aligned knots (pointing towards the apex
of the bow shock). Later observations showed the existence of a Northern
counterpart to HH 34S (HH34N) and of a series of bipolar bow shock
pairs at larger distances from the outflow source (Bally \& Devine 1994;
Eisl\"offel \& Mundt 1997; Devine et al. 1997).
There is a wealth of observations of this outflow, including images,
spectrophotometry, radial velocity and proper motion measurements at optical
(see, e. g., Heathcote \& Reipurth 1992; Eisl\"offel \& Mundt 1992;
Morse et al. 1992, 1993; Reipurth et al. 2002; Beck et al. 2007)
and IR wavelemgths (Stapelfeldt et al. 1991; Stanke et al. 1998;
Reipurth et al. 2000). These observations
show that the HH~34 outflow (at a distance of $\approx 417$~pc,
see Menten et al. 2007) has a plane of the sky velocity of
$\approx 150$~km~s$^{-1}$ and propagates at an angle of $\approx 30^\circ$
from the plane of the sky (the southern lobe being directed
towards the observer).

The flow itself is driven by HH 34 IRS, a Class I protostar
surrounded by a relatively large ($\sim 1000$ AU radius) circumstellar disk
(Stapelfeldt \& Scoville 1993; Anglada et al. 1995). The possible
coupling of the disk and outflow has motivated the search for
signatures of internal rotation within the jet as consequence
of the transfer of angular momentum from the rotating disk into
the highly collimated jet (Coffey et al. 2010).

In spite of the clear bipolar symmetry of the bow shock pairs (extending
to $\sim 1.5$ pc from the source, see Devine et al. 1997), observations
extending over $\sim 2$ decades did not detect a nothern
counterpart for the chain
of aligned knots extending $\sim 30''$ southwards from the HH 34 source.
This situation changed with the paper of Garc\'\i a L\'opez et al. (2008),
who obtained IR (1.6 and 2.1 $\mu$m) long-slit spectra in which the
emission of the northern counter-jet was finally detected. These
authors note that the emission of the counterjet has intensity peaks
at positions (i. e., distances from the source) which approximately coincide
with the knots along the southern jet.

In this paper, we present new Spitzer IRAC images of HH 34. These
images show the southern jet and northern counterjets with comparable
intensities, and with a surprising degree of symmetry. The observations
are described in section 2. In section 3 we present an image of the
central region of the HH 34 outflow, quantitatively evaluate the
degree of symmetry between the jet and the counterjet, and discuss
the implications of the results for the ejection mechanism that
has produced the outflow. The results are summarized in section 4.

\section{Observations}

The observations of HH 34 are part of our original Spitzer Space
Telescope (Werner et al. 2004) General Observer (GO) program 3315
(PI Noriega-Crespo) obtained with both the infrared camera IRAC
(Fazio et al. 2004) and the infrared photometer MIPS (Rieke et al. 2004)
in March 28, 2005. The data have been recovered from the Spitzer Legacy
Archive and the quality of the final images (Post Basic Calibrated 
Data or Post-BCD; S18.7 products)
is outstanding, so that no further processing was required.
In this study we present the IRAC observations obtained in the four channels
(1, 2, 3, 4) = (3.6, 4.5, 5.8 \& 8.0 \mum) covering a FOV of $\sim 30\times
30$\arcmin~
(the result of a 6$\times$6 array map with a 260\arcsec~stepsize) and with a 
total integration time per pixel of 30 sec. The final images are sampled with 
0.6\arcsec~per pixel, nearly one third of standard $\sim 2$\arcsec~IRAC 
angular resolution.

Figure 1 shows a three color image of HH 34 using channels 1, 2 and 3.
Like with other protostellar outflows observed with IRAC (see 
e.g. Noriega-Crespo et al. 2004; Looney, Tobin \& Kwon 2007; Tobin et al. 2007;
Ybarra \& Lada 2009) Channel 2 recovers the strongest jet emission,  
since its bandpass (4 to 5\mum) includes three relatively bright pure 
rotational H$_2$ emission lines, 0-0 S(9) 4.69, 0-0 S(10) 4.41 and 0-0 S(11) 
4.18\mum; the jet is well detected in Channel 3 as well, where another couple 
of H$_2$ lines are found (0-0 S(6) 6.11 and 0-0 S(7) 5.51\mum) 
(see De Buizer \& Vacca 2010).

\begin{figure}[h!]
\centerline{\hbox{
\includegraphics[width=260pt,height=350pt,angle=0]{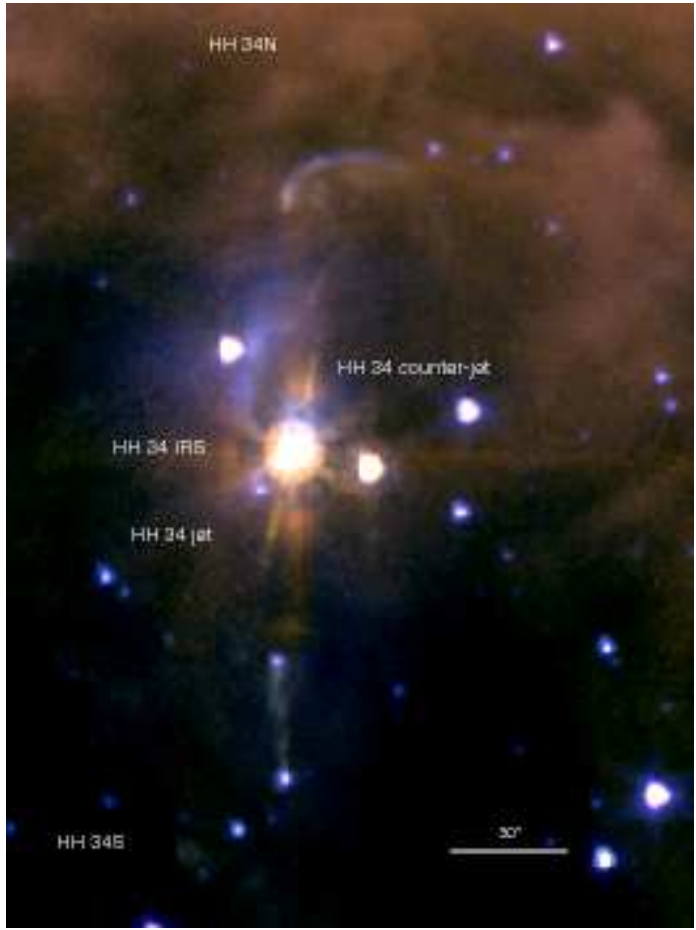}}}
\caption{A section of the HH 34 IRAC map
($\sim$ 3\arcmin $\times$ 4\arcmin) centered
on the HH 34 IRS source using the four channels
3.6 (blue) , 4.5+5.8 (green) and 8\mum~(red). North is up and
East to the left. The image shows the
counter-jet as well as a wider infrared bowshock about 20\arcsec~South
of the optical HH 34N main bow shock. Indeed both HH 34S and 34N lie
within the image, but are barely visible. Other extended
emission structures (e.~g., the North-South, jet-like
structure South of HH 34 IRS) seen in the map might
be associated with other outflows in the region.}
\label{fig1}
\end{figure}

\section{The jet/counterjet symmetry}

In Figure 2, we show the 4.5+5.8\mum\ image of the region around the
source of HH~34, rotated so that the outflow axis is parallel to
the ordinate. This image shows a surprising symmetry
between the southern jet (detected in optical images)
and the northern counterjet. We have defined a
7 pixel ($4''.2$) wide, recangular box aligned with the
HH~34 axis, within which we subtract the background
emission (defined as a linear interpolation between
the pixels immediately outside of the long edges of the
box). The result of this background emission subtraction
is shown on the right hand side of Figure 2, in which
the knot structure along the HH 34 jet and counterjet
is seen more clearly.

\begin{figure}[h!]
\centerline{\hbox{
\includegraphics[width=185pt,height=180pt,angle=0]{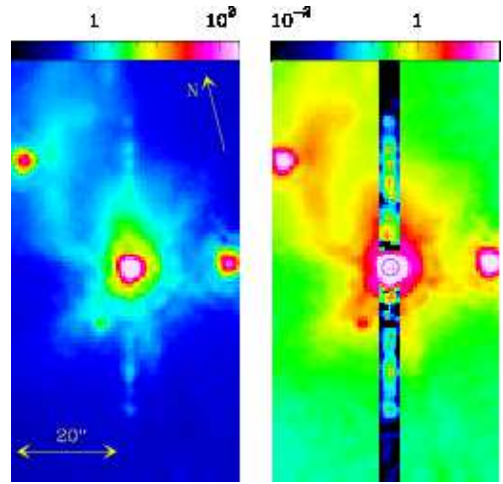}}}
\caption{Central region of the 4.5+5.8 \mum\ IRAC map (see
Figure 1), rotated so that the HH 34 axis is parallel to the
ordinate. The images correspond to the addition
of the two bands, and are displayed with the logarithmic
colour scales given (in MJy/sterad) by the bars on the top
of each plot. The left plot shows the map, and the right plot shows
a map in which a background substraction has been applied
within a rectangular box that includes the jet and the counterjet
(see section 3). The scale and orientation of the images
are shown on the left hand plot. On the right hand plot,
the red crosses show the positions of the outflow source and the
jet/counterjet knots determined with the paraboloidal fits
described in the text. The black circle shows the position
of the source calculated as the average of the positions of
the 7 jet/counterjet knot pairs (see the text).}
\label{fig2}
\end{figure}

We have carried out paraboloidal fits to the emission
peaks of the knots along the outflow axis, determining
the positions of the peaks with an error of $\approx
0.2$ pix. $= 0''.12$. Through this procedure, we
obtain the positions $x_j$ and $x_{cj}$ of the 7
knots along the jet and the counterjet (respectively).
The position of the source is not so well determined,
because the image of the source is partially saturated,
and has a complex point spread function. Because of this,
we have estimated the position of the source as the
average of the coordinates of the seven knots along
the jet and the counterjet. A paraboloidal fit to the
emission of the region around the source actually results
in a similar position (with offsets of $\approx 0.2$ pix
along and $\approx 1$ pix across the jet axis with respect
to the average position of the ensemble of knots).

In Figure 3, we plot the positions $x_{cj}$ of the consecutive
knots along the counterjet as a function of the positions
$x_j$ of the corresponding knots along the jet. This
figure shows the remarkable symmetry (with respect to
the position of the source) of the knots along the
jet and the counterjet.

\begin{figure}[h!]
\centerline{\hbox{
\includegraphics[width=180pt,height=260pt,angle=0]{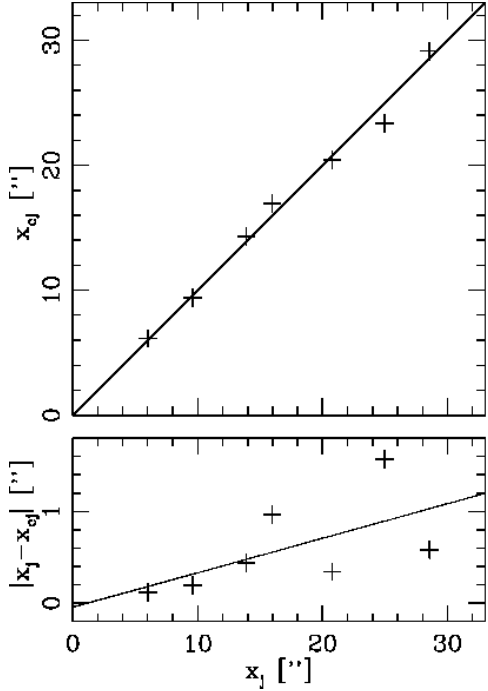}}}
\caption{The top graph shows the distance from the
source $x_{cj}$ of the knots along the northern
counterjet as a function of the positions $x_j$
of the corresponding knots along the southern jet.
The bottom plot shows the offsets $\Delta x=|x_j-x_{cj}|$
in the positions of the corresponding jet/counterjet
knots as a function of distance $x_j$ from the source.
The vertical size of the crosses in the $\Delta x$ vs. $x_j$
plot approximately correspond to the error in the observational
determinations of the $\Delta x$ values.}
\label{fig3}
\end{figure}

In Figure 3, we also plot the offsets $\Delta x=|x_j-x_{cj}|$
(between the positions of the corresponding jet/counterjet knot
pairs) as a function of $x_j$. The first three knots have
monotonically growing offsets, with $\Delta x = 0''.11\to 0''.44$
(ranging from $\sim 1\to 4$ times the measurement error, see
above). The four knots further away from the source generally have
larger offsets, with a top value $\Delta x=1''.57$ for the
6th knot out from the source. The resulting $\Delta x$ vs. $x_j$
dependence therefore has low offsets close to the source, and
increasingly large (and more highly oscillating) values for
larger distances from the source.

These results can be interpreted as follows. Let us first
assume that the jet/counterjet knot pairs are ejected at
the same time $\tau$, but with different velocities $v_j$ and
$v_{cj}=v_j-\Delta v$ (for the jet and counterjet knots,
respectively). If the knots are ballistic, at a time $t$ they
will have positions $x_j=(t-\tau)v_j$ and $x_{cj}=
(t-\tau)(v_j-\Delta v)$. One then obtains
that the jet/counterjet knot offset $\Delta x_v$ (due
to asymmetries in the jet/counterjet ejection velocities)
is given by~:
\begin{equation}
\Delta x_v=x_j-x_{cj}={\Delta v\over v_j}x_j\,.
\label{dxv}
\end{equation}
In other words, an ejection velocity asymmetry in the
simultaneous ejection of a jet/counterjet knot pair results
in a knot position asymmetry $\Delta x_v$ that increases
linearly with distance from the source (as the knots travel
away from the source).

Let us now assume that we have a pair of knots ejected
with the same velocity $v_j$, but at times $\tau$ and
$\tau+\Delta \tau$ (for the knot along the jet and the
counterjet, respectively). For ballistic knots, at a time
$t$ they would then be at distances $x_j=(t-\tau)v_j$
and $x_{cj}=(t-\tau-\Delta \tau)v_j$ from the source.
The jet/counterjet knot offset $\Delta x_\tau$ (due
to a time-defference $\Delta \tau$ in the ejection time)
is given by~:
\begin{equation}
\Delta x_\tau=x_j-x_{cj}=\Delta \tau v_j\,.
\label{dxt}
\end{equation}
In other words, the position asymmetry $\Delta x_\tau$
does not change as the knots travel away from the source
along the jet and counterjet.

For the case of jet/counterjet knot pairs ejected
with a time difference $\Delta \tau$ and a velocity
difference $\Delta v$, we therefore predict an
asymmetry between the jet and counterjet knot positions
of the form
\begin{equation}
\Delta x=|\Delta x_v + \Delta x_\tau|=
|{\Delta v\over v_j}x_j+\Delta \tau v_j|\,.
\label{dx}
\end{equation}
This relation gives the behaviour of $\Delta x$ as
a function of distance from the source $x_j$ (along
the jet) for a given pair of knots as they travel
down the jet/counterjet axis.

If we have a series of knot pairs (7 knot pairs in
our HH~34 image), the values of $\Delta v/v_j$
and $\Delta \tau v_j$ will not necessarily be identical
for all of the knot pairs. We can anyway attempt to
fit a linear $\Delta x$ vs. $x_j$ dependence to the
observations.

At large enough distances from the source, equation (\ref{dx})
is a straight line which goes through the origin (of the
$\Delta x$ vs. $x_j$ coordinates), with
a slope of $|\Delta v/v_j|$. The $\Delta \tau v_j$ term
(see equation \ref{dx}) results in a scatter of points above
and below (depending on the sign of $\Delta \tau$ this straight line.
This scatter will also have a contribution from the distribution
of the $|\Delta v/v_j|$ values over the knot pairs.

Therefore, if we have a distribution of $\Delta \tau$ (over the
different knot pairs) with a zero mean velocity (i.~e., with similar
positive and negative $\Delta \tau$ values), we would
expect to recover from the data a straight line that still
goes through the origin. The slope of this line would represent
an average of $|\Delta v/v_j|$ over the ensemble of knot
pairs (see equation \ref{dx}). The value of the determined
error for the $x_j=0$ interception point of the straight
line would give us an upper limit to
the root mean square average of $\Delta \tau v_j$.

Fitting a least squares linear dependence to the observed
offsets, we obtain
\begin{equation}
v_j\Delta \tau =(-0.04\pm 0.36)''\,;\,\,\,\,\,\,\,\,\,
{\Delta v\over v_j}=0.038\pm 0.019\,.
\label{fit}
\end{equation}
We therefore obtain an estimate for the average time offset of basically 0
(as expected), an upper limit for
its root mean square value of $0''.36/v_j$, and a velocity
asymmetry with an average modulus of approximately $(0.04\pm 0.02)v_j$.

If we consider a distance of 417~pc (to HH~34) and that the knots
have a plane of the sky velocity $v_j\approx 150$~km~s$^{-1}$,
we then obtain
\begin{equation}
\Delta \tau\leq 4.5\,{\rm yr}\,;\,\,\,\,\,\,\,\,\,
\Delta v=(5.7\pm 2.8)\,{\rm km\,s^{-1}}\,.
\label{fit2}
\end{equation}

The maximum value of $\Delta \tau$ that we have determined in
this way from the observed jet/counterjet asymmetry of HH~34
directly implies a maximum possible size for the region in which
the ejection is produced. The gas at the foot of a stellar jet
is observed to have temperatures of $\sim 10^3$~K, implying a sound
speed of $\approx 3$~km~s$^{-1}$. The Alv\'en velocity in this region
is expected to have similar values. In order to synchronize the
jet/counterjet ejections to a time $\Delta \tau$,
the region producing both outflows has to
be able to communicate causally within this time interval.
Therefore, waves propagating at
a velocity $v_s$ (this could be the Alfv\'en or the sound velocity,
both $\sim 3$~km~s$^{-1}$, see above) should be able to travel
the characteristic size $D$ of the jet formation region within
a time $\Delta \tau$. This condition can be written as~:
\begin{equation}
D\leq v_s\,\Delta \tau\,,\,\,\,
{\rm so}\,\,D\leq \left({v_s\over {\rm 3\,km\,s^{-1}}}\right)\,
2.8\,{\rm AU}\,,
\label{d}
\end{equation}
where for the second inequality we have used the value of $\Delta \tau$
derived from the observed HH~34 jet/counterjet asymmetry (see equation
\ref{fit2}) and the estimated sound speed of 3~km~s$^{-1}$.

\section{Summary and conclusions}

We have obtained IRAC images of the region around the source
of the HH~34 outflow (see Figure 1).
These images show the presence of a southern jet
and a northern counterjet (previously only detected in the IR
long-slit spectrum presented by Garc\'\i a L\'opez et al. 2008).
The seven knots along the jet and the counterjet have a remarkable
symmetry with respect to the position of the outflow source (see Figure 2).

We find that if we plot the offsets $\Delta x=|x_j-x_{cj}|$ (where
$x_j$ is the distance from the source
of a knot along the jet and $x_{cj}$ the distance to the corresponding
knot along the counterjet) as a function of the position $x_j$ of
the knots, we obtain a general trend of lager $\Delta x$ values
for increasing distances from the source (see Figure 3). This
trend can be interpreted (on the basis of a simple, ballistic
knot model) as the result of a time-offset $\Delta \tau$ (between
the ejection of the two knots in a given pair) and of a velocity
difference $\Delta v$ (between the velocities of the knot pair).

From a fit to the observed $\Delta x$ vs. $x$ relationship,
we determine a velocity asymmetry of $\Delta v=(5.7\pm 2.8)$~km~s$^{-1}$,
and an upper limit for the time-offset of the binary ejection
of $\Delta \tau=4.5$~yr. Interestingly, the velocity asymmetry
estimated for the binary ejections is within a factor of $\sim 2$
of the estimates for the sound and Alfv\'en speeds at the base
of an outflow from a young star. Actually, the estimates of the
sound and Alfv\'en speeds lie within the error bars of the
$\Delta v$ value determined from the HH 34 jet/counterjet knot
asymmetries.

It is possible to use the upper limit of $\approx 4.5$~yr to derive
an estimate $D\approx 2.8$~AU for the size of the jet production
region. For example, if one considers models of magnetocentrifugally
driven disk winds, this estimate would imply a characteristic radius
for the Alfv\'en surface (in which the outflow becomes super-Alfv\'enic)
of $D/2\approx 1.4$~AU. This is an interesting constraint on models
of the production of jets from young stars (for a recent
discussion of jet production models see the review of
Ferreira 2008, and the papers in the book of Garc\'\i a
\& Ferreira 2010).

Another possible interpretation is that the $\approx 1.4$~AU radius
(derived from the observations) corresponds to the outer radius
of the region of the accretion disk which gives rise to the jet. This
interpretation might be more comfortable in terms of magnetocentrifugally
driven disk winds, because an Alfv\'en surface of $\sim 1$~AU radius
might be too small.

VLA observations of sources of HH outflows had provided
a constraint of a maximum size of $\approx 50$~AU ($0''.1$
at the distance of Orion) on the size of the jet production
region (see, e.~g., the HH~1/2 observations of
Rodr\'\i guez et al. 2000).
The upper limit of $\approx 2.8$~AU that we have derived
above for the size of the formation region of the HH~34
jet is therefore the most stringent condition yet. This
size corresponds to $\approx 7$~mas, a resolution that
will only be approached in direct observations of an HH source
by future efforts with the ALMA interferometer. Observations
with large baseline near-IR interferometers (such as Keck-I
or VLT-I) could possibly also resolve such a region, provided
that the emission is strong enough to be detected.

To conclude, we should point out that the derived upper limit
of $\approx 2.8$~AU for the jet formation region is derived
from a simple, ballistic approximation for the motion of the
knots along the HH 34 jet and counterjet. Of course, the observed
asymmetry could partially arise from deviations from a purely
ballistic motion. If this is the case, the size implied for
the jet formation region would be even smaller.

\acknowledgements
This work is based in part on observations made with the {\it Spitzer Space 
Telescope} which is operated by the Jet Propulsion Laboratory, California 
Institute of Technology under NASA contract 1407.
The work of AR and VL was supported by the CONACyT grants 61547,
101356 and 101975.

\end{document}